\documentclass[12pt]{article}
\usepackage{graphicx}
\usepackage{cite}
\usepackage{hyperref}

\newcommand\pubnumber{UdeM-GPP-TH-12-211\\ EFI 12-19\\ TECHNION-PH-12-11\\ arXiv:1207.6390}
\newcommand\pubdate{July 26, 2012}
\def\support{\footnote{Speaker}}

\def\Title#1{\begin{center} {\Large #1 } \end{center}}
\def\Author#1{\begin{center}{ \sc #1} \end{center}}
\def\Address#1{\begin{center}{ \it #1} \end{center}}

\newcommand\pubblock{\rightline{\begin{tabular}{l} \pubnumber\\
         \pubdate  \end{tabular}}}
\newenvironment{Abstract}{\begin{quotation}  }{\end{quotation}}
\newenvironment{Presented}{\begin{quotation} \begin{center}
             PRESENTED AT\end{center}\bigskip
      \begin{center}\begin{large}}{\end{large}\end{center} \end{quotation}}

\def \b{{\cal B}}
\def \babar{B{\sc a}B{\sc ar}~}

\def \bea{\begin{eqnarray}}
\def \beq{\begin{equation}}
\def \ca{{\cal A}}
\def \cao{\overline{\ca}}
\def \cO{{\cal O}}
\def \cP{{\cal P}}
\def \cPA{{\cal PA}}
\def \cT{{\cal T}}

\def \eea{\end{eqnarray}}
\def \eeq{\end{equation}}

\def \od{\overline{D}^0}
\def \ok{\overline{K}^0}

\def \s{\sqrt{2}}
\def \st{\sqrt{3}}

\def \sx{\sqrt{6}}

\def \i{{\it i}}
\def \G{\Gamma}
\def \g{\gamma}
\def \d{\delta}
\def \ovD{\overline{D}}
\def \of{\overline{f}}
\def \db{\overline{d}}
\def \sb{\overline{s}}
\def \la{\lambda}
\def \nn{\nonumber}

\begin{document}
\begin{titlepage}
\pubblock

\vfill
\Title{NONLEPTONIC CHARM DECAYS AND CP VIOLATION}
\vfill
\Author{Bhubanjyoti Bhattacharya\support}
\Address{Physique des Particules, Universit\'e de Montr\'eal \\
C.P. 6128, succ.\ centre-ville, Montr\'eal, QC, Canada H3C 3J7}
\medskip
\Author{Michael Gronau}
\Address{Physics Department, Technion -- Israel Institute of Technology \\
Haifa 32000, Israel}
\medskip
\Author{Jonathan L. Rosner}
\Address{Enrico Fermi Institute and Department of Physics \\
University of Chicago, 5620 S. Ellis Avenue, Chicago, IL 60637}
\bigskip
\begin{Abstract}
In this talk we briefly present a flavor-SU(3) technique to study branching
ratios and direct CP asymmetries of $D$-meson decays. The first part of the talk
is meant to set up a foundation, based on previous work, to deal with flavor-SU(3)
amplitudes and relative strong phases. In addition, we present a model for dealing
with SU(3)-breaking in branching ratio measurements of SCS $D^0$ decays. In the
second part of the talk we make use of a proposal for an enhanced CP-violating
penguin in the SM, to explain the recent LHCb and CDF observations of CP violation
in SCS $D^0$ decays. Furthermore, we use our model to predict CP violation in
$\pi^0\pi^0$ and $K^+\ok$ final states. Large experimental bounds on individual CP
asymmetries give rise to a large allowed range of $\delta$, the strong phase of
the CP-violating penguin. We also briefly discuss future prospects.
\end{Abstract}
\vfill
\begin{Presented}
Charm 2012, The 5th International Workshop on Charm Physics \\
14-17 May 2012, Honolulu, Hawai'i 96822
\end{Presented}
\vfill
\end{titlepage}
\def\thefootnote{\fnsymbol{footnote}}
\setcounter{footnote}{0}


\section{Introduction: Why study charm?}

Experiments worldwide are looking for signatures of new physics beyond the
Standard Model (SM). While most searches are dedicated to energy frontiers,
if new physics is present, there is a high chance that it may rear its head
at low energies in $B$ and $D$ meson decays. In order for new physics
searches at low energies to be successful, however, it is necessary to rule
out the possibility that the observed signals may have been produced by SM
processes that are yet not understood, so that indeed the signal is new
physics. Thus, a good place to look for new physics is where the expected
SM signals are suppressed, so that it is easier to distinguish a new physics
signal from the expected SM background.

Charm physics is dominated by the first two generations of quarks and the
associated elements of the Cabibbo-Kobayashi-Maskawa (CKM) matrix in the SM.
These matrix elements are relatively large compared to those also involving
the third generation of quarks. Furthermore, since the charm mass is quite
close to the scale at which perturbative QCD breaks down ($\Lambda_{\rm QCD}
$), one expects non-perturbative effects to cause unforeseen enhancements
\cite{Golden:1989qx}. Due to these reasons one might think that nothing new
can be observed in charm physics. However, processes involving charm also
depend on the suppressed elements in the CKM matrix associated with the
third generation of quarks via CKM unitarity constraints. Utilizing this
fact, one may still be able to use charm physics as a possible source for
new physics signals, in particular in studies of CP violation in charmed
meson decays.

Earlier the CDF collaboration had measured time-integrated CP asymmetries in
$D^0$ decays to $K^+K^-$ and $\pi^+\pi^-$ final states
\cite{Aaltonen:2011se}, and found them to be consistent with zero. Recently,
however, the LHCb collaboration found 3.5 $\sigma$ evidence for direct CP
violation in singly-Cabibbo-suppressed (SCS) $D$ decays \cite{Aaij:2011in}.
LHCb measured the difference between direct CP asymmetries in the above
channels, and found it to be approaching the percent level. This has brought
charm physics into the limelight as a potential source for new physics signals.
Based on the LHCb result, CDF has presented an updated analysis
\cite{Collaboration:2012qw}. The joint LHCb and CDF results now measure
almost a $- 0.7\%$ difference in CP asymmetries in the two channels. SCS $D$
decays in particular have thus recently received an increased amount of
attention. Many authors have proposed models that involve new physics at an
appropriately high mass scale, that may be responsible for a large observed
direct CP asymmetry \cite{Grossman:2006jg, Bigi:2011re, Bigi:2011em,
Isidori:2011qw, Wang:2011uu, Rozanov:2011gj, Hochberg:2011ru,
Altmannshofer:2012ur}. These models also
include additional ways of testing their validity. Some authors, however,
have proposed a more cautious line of thought, arguing that non-perturbative
enhancements in SM amplitudes can't be ruled out and may be responsible for
the large observed CP violation \cite{Brod:2011re, Pirtskhalava:2011va,
Cheng:2012wr, Cheng:2012xb, Cheng:2012cu}. We studied the phenomenological
consequences of such an enhanced penguin in Refs.\ \cite{Bhattacharya:2012ah,
Bhattacharya:2012kq}.

We begin by briefly discussing the technique of applying flavor-SU(3)
symmetry to study charm decays. Sec.\ II presents a discussion of
branching fractions in Cabibbo-favored (CF) $D$ decays, studied in the
light of flavor-SU(3) symmetry. We discuss a model for studying
flavor-SU(3) breaking in SCS decays and also discuss $D$ decays to a
pseudoscalar (P) and a vector (V) meson. In Sec.\ III we discuss our
model based on a phenomenological CP-violating penguin amplitude,
required to explain the LHCb and CDF observations. We also discuss the
possibility of measuring similar sub-percent level CP asymmetries in
other $D^0$ and $D^+$ decay channels. We present our conclusions in
Sec.\ IV.

\section{Branching fractions in decays of $D$ mesons}

\subsection{Flavor SU(3) diagrammatics}

In this section we review the basics of flavor-SU(3) diagrammatics. Cabibbo-favored
(CF) decays are described in terms of tree-level topologies ``Color-favored Tree''
($T$), ``Color-suppressed Tree'' ($C$), ``Exchange'' ($E$) and ``Annihilation''
($A$). The quark-level transition $c\to s u \db$ does not allow penguin topologies.
In order to describe singly-Cabibbo-suppressed (SCS) $D$ decays, in addition to
tree-level topologies we also need ``Penguin'' ($P$) and ``Penguin Annihilation''
($PA$) topologies associated with the quark-level transitions $c\to d u \db$ and
$c\to s u \sb$. Furthermore, in order to account for flavor-SU(3) breaking in SCS
processes, we include factorizable SU(3)-breaking in $T$ and $A$ amplitudes, as
described originally in Ref.\ \cite{Bhattacharya:2012ah} and outlined in the
following expressions:
\bea
T_{D^0\to\pi^+\pi^-}~~=~~T_{D^+\to\pi^+\pi^0}~~=~~T_{D^+_s\to\pi^+K^0}~~=~~T_\pi~,\\
T_{D^0\to K^+K^-}~~=~~T_{D^+\to K^+\ok}~~=~~T_K~,~~~~~~~~\\
A_{D^+_s\to\pi^+K^0}~~=~~A_{D^+_s\to K^+\pi^0}~~=~~A~,~~\\
A_{D^+\to K^+\ok}~~=~~A_{D^+}~~=~~A\cdot\frac{f_{D^+}}{f_{D^+_s}}~.~~~~~~~~
\eea
where, $T_{\pi, K}$ are obtained in terms of known parameters, namely the CF $T$
\cite{Bhattacharya:2009ps}, relevant form factors \cite{Shipsey:2007, Besson:2009uv},
decay constants \cite{Rosner:2010} and meson masses \cite{Nakamura:2010zzi}, as
follows:
\bea
T_\pi &=& T\,\cdot\,\frac{|f_{+(D^0\to\pi^-)}(m^2_\pi)|}{|f_{+(D^0\to K^-)
(m^2_\pi)|}} \,\cdot\, \frac{m^2_D - m^2_\pi}{m^2_D - m^2_K}~, \\
T_K   &=& T\,\cdot\,\frac{|f_{+(D^0\to  K^-)}(m^2_K)|}{|f_{+(D^0\to K^-)}
(m^2_\pi)|} \,\cdot\,\frac{f_K}{f_\pi}.
\eea

Tree-level topologies in CF-decay amplitudes are proportional to a CKM factor
$V^*_{cs}V_{ud}\sim 1$, while those in SCS-decay amplitudes come with a factor
of $V^*_{cd}V_{ud}\sim -\la$ or $V^*_{cs}V_{us}\sim\la$ (we neglect the weak-
phase difference between these quantities that shows up at a higher order in
$\la$). Here we use $\la = \tan\theta_C = 0.2317$ \cite{Nakamura:2010zzi}.

Each relevant penguin amplitude gets contributions from all three down-type quarks
running in the loop. Therefore, unlike the tree topologies, penguin topologies
depend on more than one CKM factors. However, unitarity of the CKM matrix tells
us that these CKM factors are not completely independent, but that they obey the
relationship:
\beq
V^*_{cd}V_{ud} + V^*_{cd}V_{ud} + V^*_{cb}V_{ub} = 0~.\\
\eeq
Using CKM unitarity, thus, it is possible to eliminate one of the CKM factors
as follows:
\bea
\sum_q V^*_{cq}V_{uq} \cP_q &=& V^*_{cs} V_{us} (\cP_s - \cP_d) + V^*_{cb} V_{ub}
(\cP_b - \cP_d)~,\\
&=& P + P_b~,
\eea
where the index $q$ denotes the quark running in the loop, and $\cP$ denotes the
reduced matrix element corresponding to the penguin topology denoted by $P$. Thus,
in the last line we have used the following definitions of the variables:
\bea
V^*_{cs} V_{us} (\cP_s - \cP_d) &\equiv& P~,\\
V^*_{cb} V_{ub} (\cP_b - \cP_d) &\equiv& P_b~. \label{eqn:rdc}
\eea
We now note that the magnitude of $P_b$ is largely suppressed due to the CKM
factor $V^*_{cb}V_{ub} \sim \cO(\la^5)$ when compared to the magnitude of $P$.
Thus, in a discussion solely concerning branching fractions of $D$ decays it
is justified to neglect $P_b$. However, $P_b$ is relevant in the discussion
of direct CP asymmetries, as we shall see later.

\subsection{$D\to PP$ decays}

CF $D$ decays to a pair of pseudoscalars were examined in detail in Refs.\
\cite{Bhattacharya:2009ps, Bhattacharya:2007jc, Bhattacharya:2008ss,
Cheng:2010ry}. In short, there are eight decay rates that depend on four
complex amplitudes and three relative strong phases. In Ref.\
\cite{Bhattacharya:2009ps}, a $\chi^2$-minimization fit was used to obtain
these parameters. The ($|T| > |C|$) solution obtained in this fit is:
\bea
T &=& 2.927~, \\
C &=& 2.337\,e^{-\,\i\,151.66^\circ} = -2.057 - 1.109~i~, \label{C}\\
E &=& 1.573\,e^{~\i\,120.56^\circ} = -0.800 + 1.355~i~, \\
A &=& 0.33\,e^{~\i\,70.47^\circ} = 0.110 + 0.311~i~\label{E},
\eea
in units of $10^{-6}$ GeV. The minimum value for $\chi^2$ was found to be
$1.79$ for one degree of freedom, indicating that the fit was reasonable.
In Table \ref{tab:CF2P} we list the CF decay modes and their flavor-topology
representations, alongside the observed and fitted branching fractions for
comparison.
\begin{table}
\caption{Representations and comparison of experimental and fit
branching fractions for CF decays of charmed mesons to two pseudoscalars
\cite{Bhattacharya:2009ps}. \label{tab:CF2P}}
\begin{center}
\begin{tabular}{|c l c c c|}
\hline \hline
Meson & Mode      & Rep.($\ca$)& ${\b}$ ($\%$)  & Fit ${\b}$ ($\%$) \\ \hline
$D^0$ &$K^- \pi^+$ & $T+E$&3.89$\pm$0.08 & 3.91 \\
       &$\ok\pi^0$ & $(C-E)/\s$&2.38$\pm$0.09 & 2.35 \\
       &$\ok \eta$ & $C/\st$&0.96$\pm$0.06    & 1.00 \\
       &$\ok \eta'$& $-(C+3E)/\sx$&1.90$\pm$0.11 & 1.92 \\ \hline
$D^+$  &$\ok \pi^+$& $C+T$&3.07$\pm$0.10 & 3.09 \\ \hline
$D^+_s$&$\ok K^+$  & $C+A$&2.98$\pm$0.17 & 2.94 \\
       &$\pi^+\eta$&$(T-2A)/\st$&1.84$\pm$0.15& 1.81 \\
       &$\pi^+\eta'$&$2(T+A)/\sx$&3.95$\pm$0.34& 3.60 \\ \hline \hline
\end{tabular}
\end{center}
\end{table}

Although the flavor-SU(3) parameters fit quite reasonably the decay rates
of CF $D$ decays to a pair of pseudoscalars, in order to describe SCS decays,
however, one needs to include the effects of flavor-SU(3) breaking. The
necessity to include SU(3)-breaking effects can be illustrated by considering
the SCS processes $D^0\to\pi^+\pi^-, K^+ K^-, K^0\ok$. Although under flavor
SU(3) the amplitudes for the first two processes are the expected to be
identical, in practice both these amplitudes differ from their predicted
value. Furthermore the amplitude for the third processes is predicted to
vanish under flavor SU(3), yet in practice it is significantly different from
zero.

As a qualitative explanation to SU(3)-breaking in $D^0\to\pi^+\pi^-, K^+ K^-$,
ratios of form factors and decay constants have been used, since the amplitudes
for these processes are largely dominated by $T$ which is expected to follow
factorization. However, as discussed in Ref.\ \cite{Bhattacharya:2012ah},
corrections to $T$ using factorization fail to explain the discrepancy between
theory and experiment. This can be remedied by adding the ordinarily GIM-%
suppressed \cite{Glashow:1970gm} $P$ and $PA$ topologies, the magnitude and
strong phase of which may then be obtained by fitting to the SCS $D$-decay rates.

\begin{table}
\caption{Representations and comparison of experimental and fit
amplitudes for SCS decays of charmed mesons to two pseudoscalar mesons.
Also listed are individual $\chi^2$ contribution and overall strong phase
($\phi^f_T$) for each process \cite{Bhattacharya:2012ah}.
\label{tab:SCS2P}}
\begin{center}
\begin{tabular}{|c c c c c c|} \hline \hline
Decay & Amplitude      & \multicolumn{2}{c}{$|\ca|$ ($10^{-7}$ GeV)} & $\chi^2$
& $\phi^f_T$\\ \cline{3-4}
Mode & representation & Experiment & Theory & & degrees\\ \hline
$D^0\to\pi^+\pi^-$ &$-\lambda\,(T_\pi + E) + (P + PA)$  &4.70$\pm$0.08&4.70&0&--158.5\\
$D^0\to  K^+  K^-$ &~$\lambda\,(T_K + E) + (P + PA)$ &8.49$\pm$0.10&8.48&0.01&32.5\\
$D^0\to\pi^0\pi^0$ &$-\lambda\,(C - E)/\s - (P + PA)/\s$&3.51$\pm$0.11&3.51&0&60.0\\ \hline
$D^+\to\pi^+\pi^0$ &$-\lambda\,(T_\pi + C)/\s$ &2.66$\pm$0.07&2.26&33&126.3\\ \hline
$D^0\to K^0\ok$    &$-(P + PA) + P$                 &2.39$\pm$0.14&2.37&0.02&--145.6\\
$D^+\to K^+\ok$    &~$\lambda\,(T_K - A_{D^+}) + P$    &6.55$\pm$0.12&6.87&7&--4.2\\
$D^+_s\to\pi^+ K^0$&$-\lambda\,(T_\pi - A) + P$        &5.94$\pm$0.32&7.96&40&174.3\\
$D^+_s\to\pi^0 K^+$&$-\lambda\,(C + A)/\s - P/\s$      &2.94$\pm$0.55&4.44&7&16.4\\
\hline \hline
\end{tabular}
\end{center}
\end{table}

In Table \ref{tab:SCS2P}, we present the SCS decay modes and their flavor%
-topology representations, alongside the observed and fitted amplitudes
for comparison. Also listed in Table \ref{tab:SCS2P} is the overall strong
phase ($\phi^f_T$) for each final state $f$. (The index $T$ refers to the
CP-conserving nature of the amplitude, that is, it has the same weak phase
as the tree-level contributions.) The decay rates for $D^0\to(\pi^+\pi^-,
\pi^0 \pi^0, K^+K^-)$ depend only on the combination $P + PA$, while those
for $D^0\to K^0\ok, D^+\to K^+\ok, D^+_s\to\pi^+ K^0, \pi^0 K^+$ depend
only on $P$. We perform $\chi^2$ minimization fits and obtain the following
results for $P + PA$ and $P$:
\bea
P + PA &=& [(0.44\pm0.23) + (1.41\pm0.36)~i] \times 10^{-7}~{\rm GeV}~; \\
\chi^2/{\rm d.o.f.} &=& 0.012/1 = 0.012~.\\
P &=& [(-1.52\pm0.15) + (0.08^{+0.38}_{-0.32})~i] \times 10^{-7}~{\rm GeV}~; \\
\chi^2/{\rm d.o.f.} &=& 54/2 = 27~.
\eea
The $\chi^2$ contribution for each process has also been listed in Table
\ref{tab:SCS2P}.

\begin{figure}
\begin{center}
\includegraphics[width=0.48\textwidth]{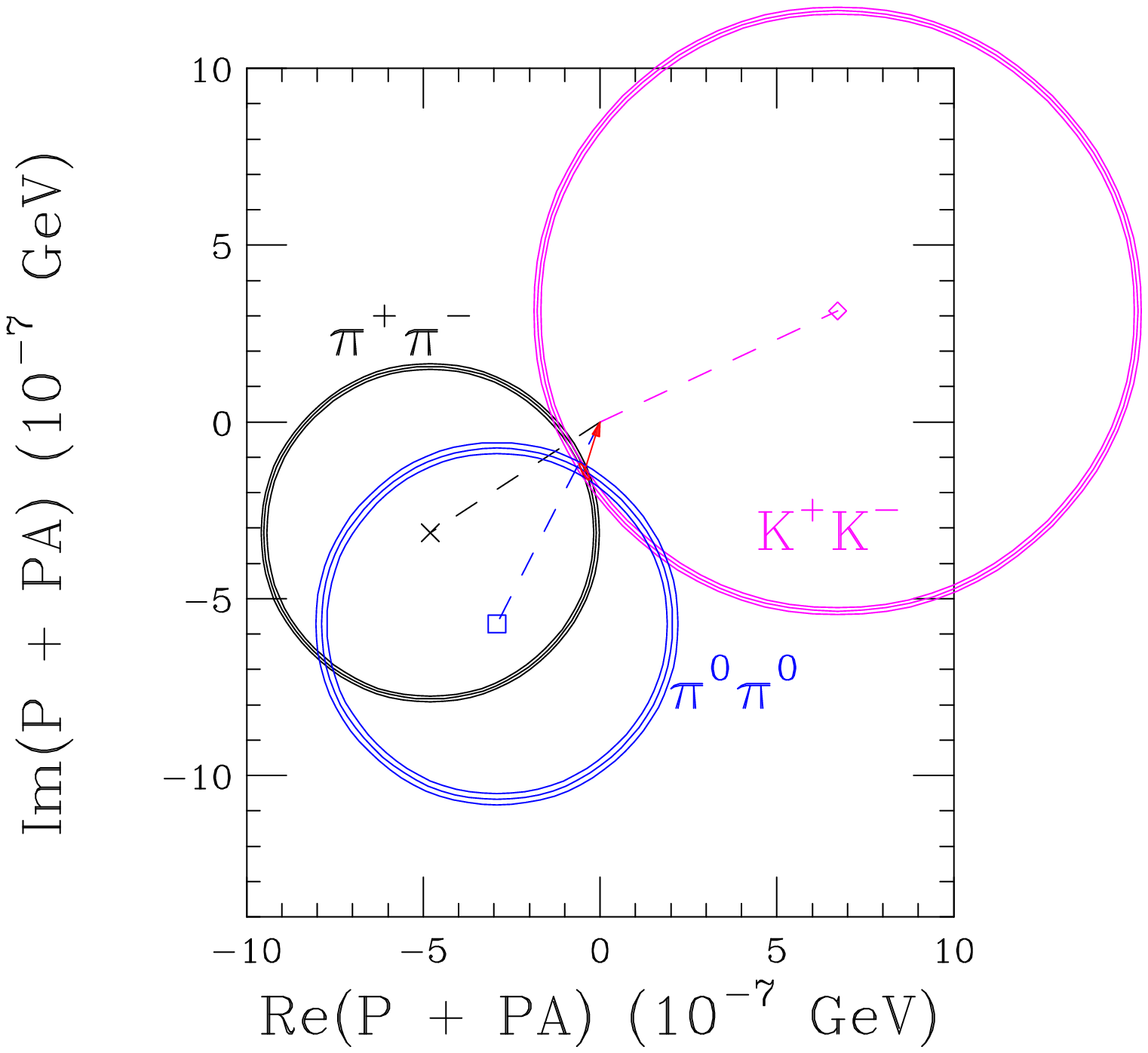} \hspace{0.4cm}
\includegraphics[width=0.42\textwidth]{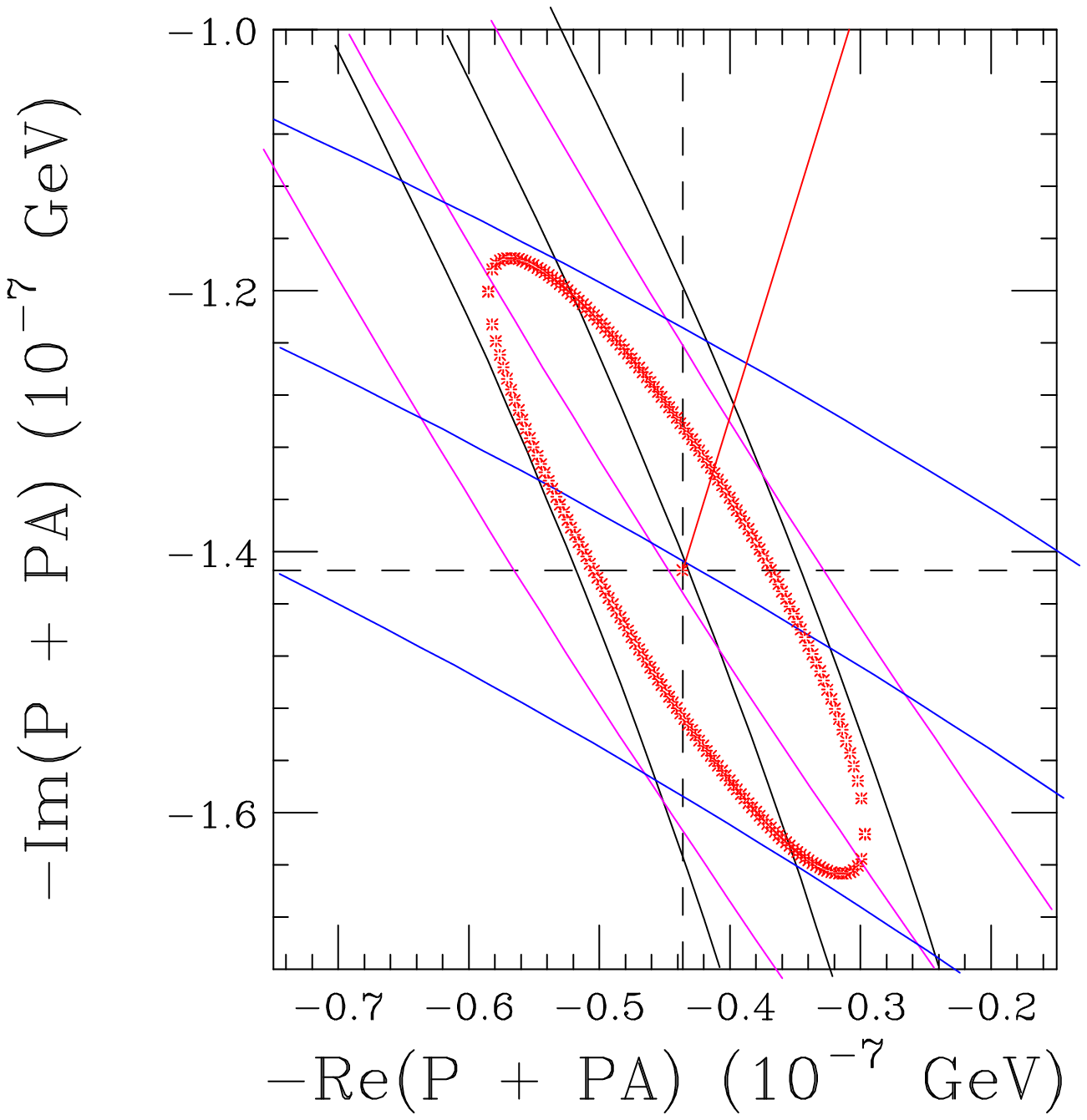}
\caption{Construction to determine $P + PA$.
The relative sign between the left-hand and (magnified) right-hand
panels is due to the fact that the vector $P + PA$ points {\it toward}
the origin in the left-hand figure.
 \label{fig:pen}}
\end{center}
\end{figure}

\begin{figure}
\begin{center}
\includegraphics[width=0.5\textwidth]{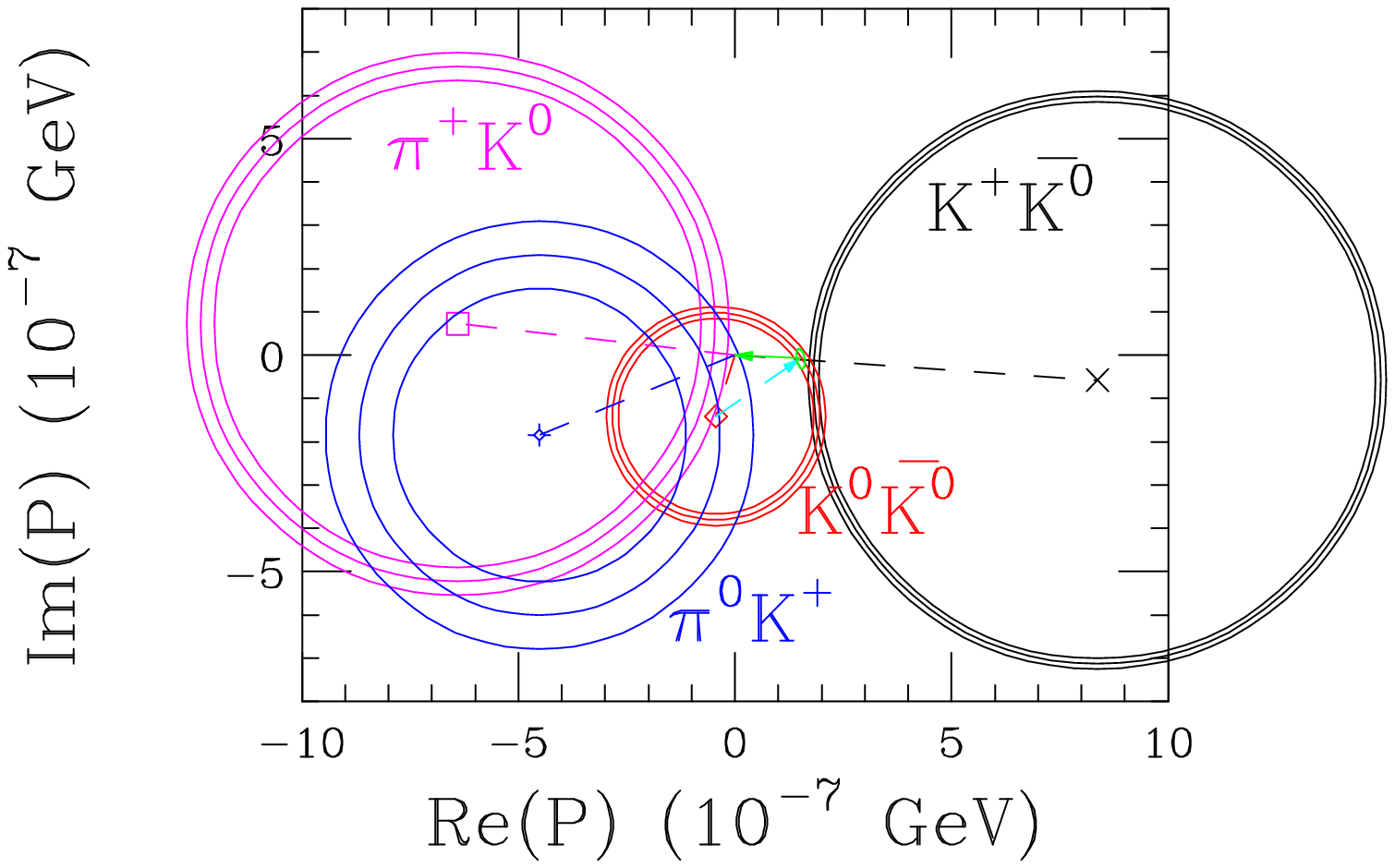} \hspace{0.2cm}
\includegraphics[width=0.44\textwidth]{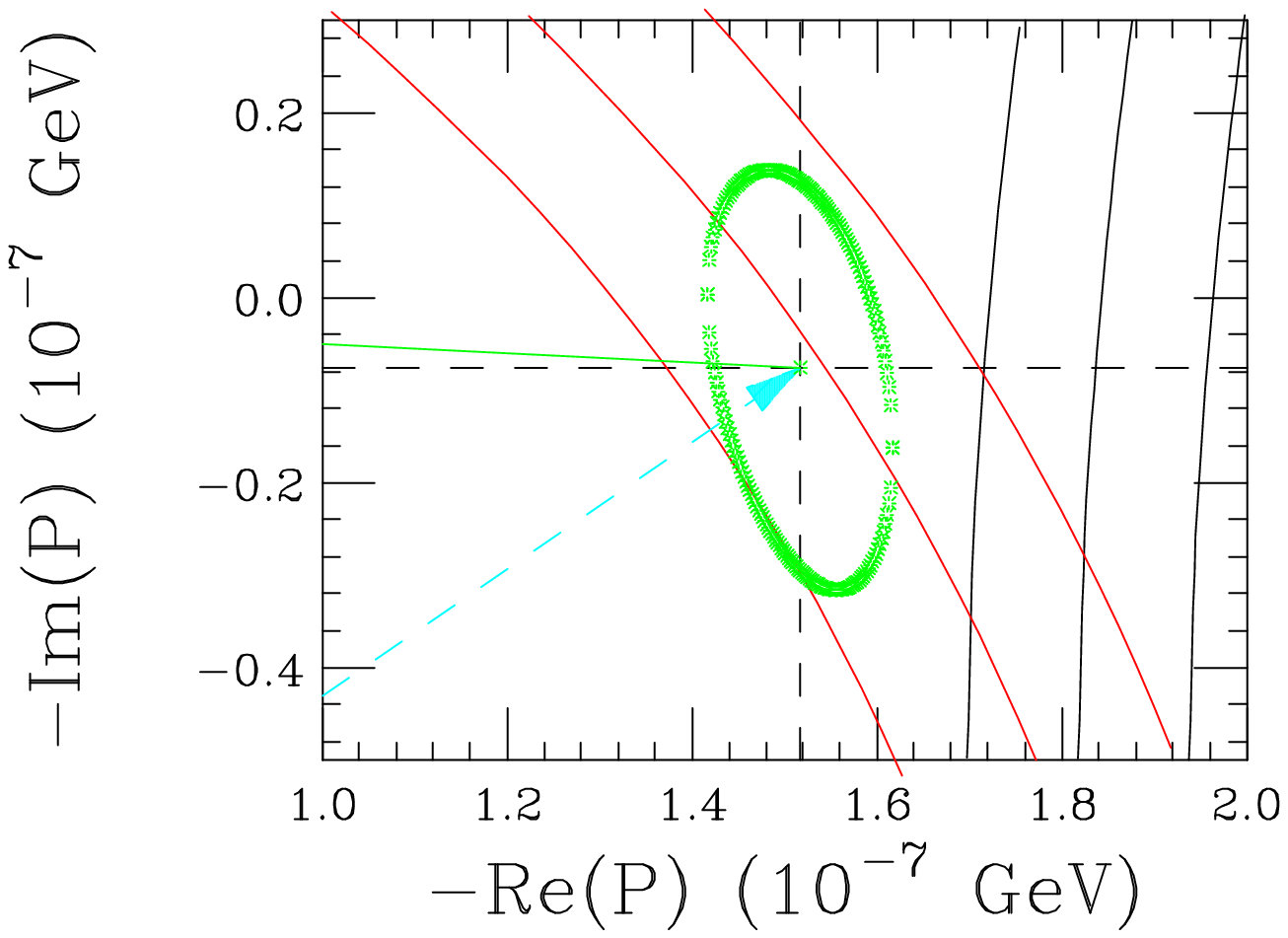}
\caption{Construction to determine $P$.
The relative sign between the left-hand and (magnified) right-hand
panels is due to the fact that the vector $P$ points {\it toward}
the origin in the left-hand figure.
 \label{fig:pe}}
\end{center}
\end{figure}

The $\chi^2$ minimum solution obtained for $P + PA$ above supports the
existence of a self-consistent solution in terms of a construction
technique described in \cite{Bhattacharya:2012ah}. We briefly revisit
the idea. Let us consider the three decays that depend only on the
combination $P + PA$. The representations for these three amplitudes
have been listed in the first three rows of Table \ref{tab:SCS2P}.
One can rewrite the representations such that the coefficient of
$P + PA$ is always one. One part of these amplitudes can be calculated
from the CF-decay parameters (including factorizable SU(3) breaking in
the $T$ diagram). We first construct vectors on the complex plane to
represent these. With the heads of these vectors as the centers we draw
circles with radii equal to the respective measured amplitudes and
their $\pm 1 \sigma$ error bands. We identify the best-possible
intersection point of the circles representing the three different
processes. The vector joining this point to the origin is then the
best solution obtained using the construction technique. A similar
exercise when performed using the last four amplitudes in Table
\ref{tab:SCS2P} gives a solution for $P$. (Note that the only unknown
quantity here is $P$, since $P + PA$ has already been determined
independently.) Fig.\ \ref{fig:pen} shows the construction technique
for obtaining $P + PA$, while Fig.\ \ref{fig:pe} shows it for $P$.

The minimum $\chi^2$ obtained for the fit to extract $P$ was
found to be much larger than what is expected for a fair fit.
In Fig.\ \ref{fig:pe}, the construction technique shows that
there is no clear region of overlap of the circles implying
that we do not have a self-consistent solution, as we do
in the case of $P + PA$. However, the extraction method for
$P$ uses $D^+_s$ decay-rates. Although one reason for not
finding a self-consistent solution for $P$ could be the
need to find a different parametrization of SU(3) breaking
in these decays, since the relevant $D^+_s$ decay rates have
large experimental errors it is difficult to study SU(3)
breaking using these decays.

Finally we note that in this section we have completely neglected
all effects of CP violation. This was justified on the basis that
for amplitude studies, the highly CKM-suppressed term $P_b$ may
be ignored. Notice that the $P + PA$ terms are proportional to
$V^*_{cs}V_{us}$. Although in cases such as $D^0\to K^+K^-$ this
term has the same weak-phase as the relevant tree-level topology
(in which case there is no CP asymmetry in the absence of $P_b$),
this is not true for example in the case of $D^0\to\pi^+\pi^-$
where the tree-level diagrams are proportional to $V^*_{cd}V_{ud}$.
The tiny weak-phase difference between $V^*_{cd}V_{ud}$ and
$V^*_{cs}V_{us}$ should give rise to a non-zero CP asymmetry in
such cases. Toward the end of Sec.\ IIIB, however, we
shall argue that a direct CP asymmetry arising from this tiny
phase difference is at least an order of magnitude smaller than
the observed CP asymmetries and hence may be neglected since the
relevant experimental error bars are currently much larger than ten
percent.

\subsection{$D\to PV$ decays}

Flavor-SU(3) techniques are also very useful in studying $D\to PV$
processes. The related diagrammatics are slightly more complicated
since one has to keep track of whether the spectator quark ends up
in the P or the V final state. The number of flavor-SU(3) parameters
also increases two-fold. Thus in place of $T$ in $D\to PP$, we now
have $T_P$ and $T_V$ in $D\to PV$, where the subscript refers to the
final state in which the spectator quark ended up. Fortunately, there
are also many more observable branching fractions. $D\to PV$ processes
were studied in great detail in Refs.\ \cite{Cheng:2010ry,
Bhattacharya:2008ke}. It was found in Ref.\ \cite{Bhattacharya:2008ke}
that under simple assumptions there are 12 discretely different
solutions for $T_P$, $C_V$, $E_V$, $T_V$, $C_P$ and $E_P$ that fit the
observed branching fractions for CF $D\to PV$ decays. The discrete
ambiguity was resolved by applying these parameters to study SCS
$D\to PV$ decays, thus allowing for the choice of a preferred $\chi^2$
minimum solution.

The flavor-SU(3) parameters extracted from $D\to PV$ may then be put to
test in three-body decays, where the final state has three pseudoscalars,
two of which are obtained as a result of decay from an intermediate
vector resonance. The Dalitz plots for several $D^0\to PV\to 3P$ processes
were studied, and several notable features supporting the validity of a
flavor-SU(3) approach were found in Refs.\ \cite{Bhattacharya:2009ps}.
Here we mention very briefly one of the striking successes of this
technique when applied to the $D^0\to\pi^+\pi^-\pi^0$ Dalitz plot.

\begin{figure}
\begin{center}
\includegraphics[width=0.55\textwidth]{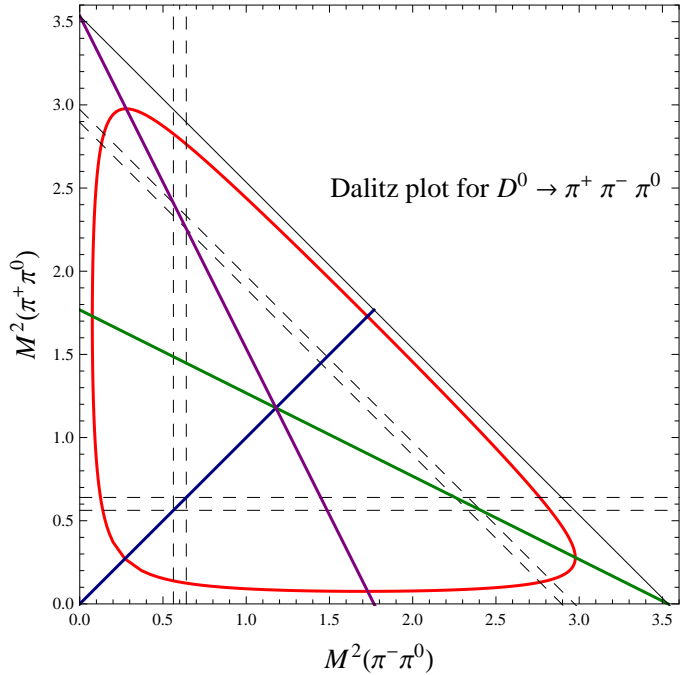}
\caption{Kinematically allowed region in Dalitz plot for $D^0\to\pi^+\pi^-
\pi^0$. Also shown are the symmetry axes in blue, green and purple. Bands
between dashed lines represent expected $\rho$ resonance bands.
\label{fig:dp}}
\end{center}
\end{figure}

Although the $\pi^+\pi^-\pi^0$ final state gets contributions from $I = $
0, 1, and 2 amplitudes, \babar found that the $I = 0$ channel dominates the
decay process $D^0\to\pi^+\pi^-\pi^0$. This conclusion can be reached at
simply by looking at the Dalitz plot for the said process. The pions have
$I = 1$. The only way to form an $I = 0$ combination of three pions is
if the wave function were totally antisymmetric under the interchange of
any two pions. On the Dalitz plot this implies that the wave function for
an $I = 0$ state must vanish along the symmetry axes shown in Fig.\
\ref{fig:dp}. \babar saw very strong depopulation along the symmetry axes
of the Dalitz plot and an isospin analysis confirmed the $I = 0$ dominance
\cite{Gaspero:2008rs}.

A flavor-SU(3) analysis \cite{Bhattacharya:2010id} in terms of $D\to PV$
amplitudes supports this observation. The relative strong phases between
interfering resonance amplitudes obtained through flavor SU(3) allow for
almost complete cancellation of the $I = $1 and 2 amplitudes, leading to
an $I = 0$ dominance. In Table \ref{tab:D3Pi} we present a comparison of
theory and experiment on different isospin amplitude contributions. This
example shows that relative strong phases between amplitudes $D$ decays
are obtained fairly successfully using a flavor-SU(3) analysis. Studies
of several other Dalitz plots \cite{Bhattacharya:2010ji,
Bhattacharya:2010tg, Bhattacharya:2012fz} show that this technique may
also be useful in comparing relative phases between interfering vector
resonances and cross-ratios between amplitudes in multiple Dalitz plots.
A more detailed discussion of relative phases in $D^0\to 3P$ processes
may be found in \cite{Bhattacharya:2011}.

\begin{table}
\caption{Comparison of theory vs.\ experiment on different isospin amplitude
contributions to the $D^0\to\pi^+\pi^-\pi^0$ Dalitz plot. \label{tab:D3Pi}}
\begin{center}
\begin{tabular}{|c c c|} \hline \hline
   Channel & Fraction($\%$) \cite{Bhattacharya:2010id} & vs \babar($\%$)
                                              \cite{Gaspero:2008rs}  \\ \hline
   I = 0 & 92.9$\pm$6.7 & 94.24$\pm$0.40 \\
   I = 1 & 4.8$\pm$0.3 & 2.17$\pm$0.17 \\
   I = 2 & 2.3$\pm$0.8 & 3.58$\pm$0.29 \\ \hline \hline
\end{tabular}
\end{center}
\end{table}

\section{Direct CP asymmetries in $D\to 2P$}

\subsection{Recent measurements from CDF and LHCb}

In our discussion so far we have neglected the CKM-suppressed
amplitude $P_b$. The addition of $P_b$ is expected to have
little-to-no effect in case of branching fractions. In this
section, however, we emphasize that this amplitude cannot be
completely neglected. $P_b$ may become particularly important
in the discussion of direct CP asymmetries in $D$ decays.

The time-integrated CP asymmetry in the process $D^0\to f$ is
defined as follows:
\beq
A_{CP} \equiv \frac{\G(D^0\to f) - \G(\od\to\of)}{\G(D^0\to f)
+ \G(\od\to\of)}
\eeq
In the special case that the final state $f$ is a CP eigenstate
(such as $\pi^+\pi^-, K^+K^-$), one can express the above as a
sum of two terms, direct CP asymmetry for the decay and an
indirect CP asymmetry associated with $D^0 - \od$ mixing. The
time-integrated CP asymmetry may then be expressed as follows:
\beq
A_{CP} \approx A^{\rm dir}_{CP} + \frac{\langle t\rangle}{\tau_D}
A^{\rm ind}_{CP}~,
\eeq
where $\langle t \rangle$ is the average decay time in the sample used and
$\tau_D$ is the true lifetime of the $D$ meson. The indirect
CP asymmetry $A^{\rm ind}_{CP}$ is known to be universal to
a very good approximation. Thus the difference between the CP
asymmetries in $D^0\to K^+K^-$ and $D^0\to\pi^+\pi^-$ is
\bea
\Delta A_{CP} &\equiv& A_{CP}(K^+K^-) - A_{CP}(\pi^+\pi^-) \nn \\
&=& \Delta A^{\rm dir}_{CP} + \frac{\Delta\langle t\rangle}{\tau_d}
A^{\rm ind}_{CP}~.
\eea
In the small $\Delta\langle t\rangle$ limit, $\Delta A_{CP}$
simply measures the difference in direct CP asymmetries in the two
channels $D^0\to K^+K^-$ and $D^0\to\pi^+\pi^-$.

CP violation in $D$ decays is expected to be small within the
Standard Model, owing to the large Cabibbo suppression of the
relevant CP-violating amplitudes. The CDF collaboration
measured the time-integrated CP violation in $D^0\to(\pi^+
\pi^-, K^+K^-)$ to be \cite{Aaltonen:2011se}:
\beq
A_{CP}(D^0 \to K^+ K^-) = (-0.24 \pm 0.22 \pm 0.09)\%,~
A_{CP}(D^0 \to \pi^+ \pi^-) = (0.22 \pm 0.24 \pm 0.11)\%~.
\eeq
We find that the corresponding 90\% confidence level limits are:
\beq \label{eqn:limits}
-0.63\% \le A_{CP}(D^0 \to K^+ K^-) \le 0.15\%~,~~
-0.21\% \le A_{CP}(D^0 \to \pi^+ \pi^-) \le 0.65\%~,
\eeq
which are consistent with a no-CP violation hypothesis. However,
a recent LHCb measurement \cite{Aaij:2011in} found $3.5 \sigma$
evidence for direct CP violation in $D^0$ decays by measuring
the difference between CP asymmetries in the two processes:
\beq \label{eqn:LHCb}
\Delta A_{CP} \equiv A_{CP}(K^+ K^-) - A_{CP}(\pi^+ \pi^-) =
[-0.82 \pm 0.21({\rm stat}) \pm 0.11({\rm syst})]\%~.
\eeq
This result was found to be consistent with the individual time-%
dependent asymmetry measurements from CDF. By knowing the average
decay time in samples for both processes, the CDF collaboration
reanalysed their data and also found evidence for direct CP
violation at a similar, fraction-of-a-percent level.  (The indirect
CP asymmetry is found to be of the order of $10^{-4}$, so that for
small decay time acceptances, the time-integrated asymmetry is
close to the direct CP asymmetry.) The combination of the CDF and
LHCb results, assuming fully uncorrelated uncertainties, was
obtained by the CDF collaboration \cite{Collaboration:2012qw}:
\beq
\Delta A^{\rm dir}_{CP}~=~(-0.67\pm0.16)\%~;~~~\Delta A^{\rm ind}
_{CP}~=~(-0.02\pm0.22)\%~. \label{eqn:DAav}
\eeq
The CDF and LHCb results are at least an order of magnitude larger
than the natural SM prediction. This has motivated many authors to
offer methods of reconciliation both within the SM as well as beyond.

Although the SM is believed to naturally predict a small value of
$\Delta A^{\rm dir}_{CP}$, an order of magnitude enhancement in hadronic
matrix elements associated with $P_b$ is not unlikely. In fact, that non%
-perturbative effects may enhance penguin amplitudes in SCS $D$ decays,
similar to the observed $\Delta I = 1/2$ enhancement in $K\to\pi\pi$,
was pointed out by Golden and Grinstein more than two decades ago in
Ref.\ \cite{Golden:1989qx}. This is not surprising, since we have
evidence for enhancement in the exchange amplitude which is expected
to be formally power-suppressed by the mass of the charm quark ($m_c$).
The charm quark mass is not far above $\Lambda_{\rm QCD}$. Thus there can
be large corrections to such formally power-suppressed terms. A full
non-perturbative calculation to extract $P_b$, however, is extremely
difficult and is often associated with sizable uncertainties.

Assuming such enhancements are possible, in the following section, we
phenomenologically constrain the magnitude and strong phase of the CP-%
violating penguin using the CDF and LHCb measurements. Using the
extracted magnitude and strong phase of the penguin $P_b$ we then
predict direct CP asymmetries in other $D^0$ and $D^+$ decays to two
pseudoscalars.

\subsection{Direct CP asymmetries using flavor SU(3)}

The amplitude for a general SCS $D$ decay to a final state $f$ may be
expressed in the form:
\bea
\ca(D\to f) &=& T_f + P_b \nn\\
&=&|T_f|e^{\i\phi^f_T}\left(1 + \frac{|P_b|}{|T_f|}e^{\i(\d- \phi^f_T
- \g)}\right)~, \label{eqn:A}
\eea
where $|T_f|$ and $\phi^f_T$ are respectively the magnitude and
strong phase of the CP-conserving part of the amplitude, while $\d$
and $- \g$ are respectively the strong and weak phases of the CP-%
violating term $P_b$, which we have written as $P_b = |P_b|e^{\i(\d
- \g)}$. We have assumed that the weak-phase difference between the
CP-violating term (proportional to $V^*_{cb}V_{ub}$) and the CP-
conserving term (proportional to $V^*_{cs}V_{us}$) is -- $\g$, where
$\g = 77^\circ$ is the angle in the standard CKM unitarity triangle.
(The deviation of the weak-phase difference between $V^*_{cb}V_{ub}$
and $V^*_{cs}V_{us}$ from -- $\g$ consists of $\cO(\la^4)$ terms and
may be neglected compared to $\g$). In writing Eq.\ (\ref{eqn:A}),
we have also assumed that a $PA_b$ contribution to the relevant CP
asymmetries may be neglected compared to the $P_b$ contribution.

In order to obtain the CP-conjugate amplitude $\cao(\ovD\to f)$ one
simply changes the sign of the weak phase, which is done by changing 
-- $\g$ to $\g$. We may now construct the direct CP asymmetries for
$D\to f$ as follows:
\bea
A_{CP}(f) &=& \frac{|\ca|^2 - |\cao|^2}{|\ca|^2 + |\cao|^2} \nn \\
&=& \frac{2(|P_b|/|T_f|)\sin\g\sin(\d - \phi^f_T)}{1 + (|P_b|/|T_f|)^2
+ 2(|P_b|/|T_f|)\cos\g\cos(\d - \phi^f_T)} \nn \\
&\approx& 2(|P_b|/|T_f|)\sin\g\sin(\d - \phi^f_T)~,\label{eqn:CPV}
\eea
where in the final line we have neglected higher order terms in $|P_b|/
|T_f|$, since this quantity is much smaller than one. Notice that the
overall strong phase of the CP-conserving term can only be obtained from
the fits up to a common sign ambiguity. However, the CP asymmetries are
approximately invariant under the joint transformations $\phi^f_T\to -
\phi^f_T$ and $\d\to\pi - \d$ as long as $|P_b|/|T_f|$ is small compared
to one. The joint LHCb -- CDF measurement of $\Delta A^{\rm dir}_{CP}$
(\ref{eqn:DAav}) may now be expressed as a constraint involving two
unknown parameters, namely $|P_b|$ and $\delta$. In addition the CDF
measurements (\ref{eqn:limits}) for the individual asymmetries may be
used to obtain upper and lower bounds on $\d$.

\begin{figure}
\begin{center}
\includegraphics[width=0.7\textwidth]{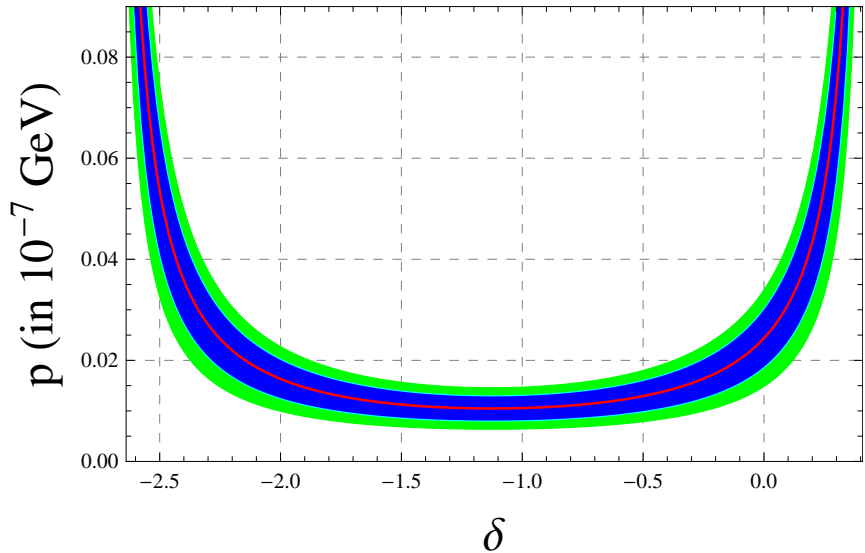}
\caption{$p \equiv |P_b|$ and $\delta$ allowed by the measured range of
$\Delta A_{CP}$. The (red) line represents the central value, while inner
(blue) and outer (green) bands respectively represent 68\% confidence level
(1$\sigma$) and 90\% confidence level (1.64$\sigma$) regions based on error
in $\Delta A_{CP}$.
\label{fig:pb}}
\end{center}
\end{figure}

In Fig.\ \ref{fig:pb} we plot the allowed range of $p\equiv|P_b|$ (up to
90\%  confidence level regions consistent with Eq.\ (\ref{eqn:DAav}) as a
function of $\delta$ in the range $-2.64 \leq \d \leq 0.41$ determined by
the CDF upper and lower bounds on the individual CP asymmetries in $D^0\to
K^+K^-$ and $D^0\to\pi^+\pi^-$ (\ref{eqn:limits}). For a large range of
$\d$, we find that $|P_b| \sim 0.01 \times 10^{-7}$ GeV, thus indicating
that the ratio $|P_b|/|T_f| \sim 1.2\times10^{-3})$ is indeed small
compared to one. (We have used $|T_f| \approx |\ca(D^0\to K^+ K^-)| = 8.46
\times 10^{-7}$ GeV.) It is interesting also to look at the ratio of the
reduced matrix elements of tree and penguin topologies (defined similarly
to Eq.\ (\ref{eqn:rdc})) which may be obtained as follows:
\bea
\frac{|\cP_b|}{|\cT_f|} &=& \frac{|V^*_{cs}V_{us}|}{|V^*_{cb}V_{ub}|}\frac
{|P_b|}{|T_f|}\nn \\
&\approx& 1.5\times 10^{3}~\frac{|P_b|}{|T_f|} \nn\\
&\sim& 2~,
\eea
where we have used Particle Data Group values for the CKM matrix elements.

Let us now revisit the question raised at the end of Sec.\ IIB. In our
conventions, the weak phase difference between $V^*_{cd}V_{ud}$ and
$V^*_{cs}V_{us}$ differs from $\pi$ by a tiny amount, which may give rise
to a non-zero direct CP asymmetry in the decay $D^0\to\pi^+\pi^-$, even in
the absence of the CP-violating penguin $P_b$. This decay amplitude may be
expressed as follows:
\bea
\ca(D^0\to\pi^+\pi^-) &=& V^*_{cd}V_{ud}~(T_\pi + E) + (P + PA) \nn \\
&=& V^*_{cd}V_{ud}~(T_\pi + E) + V^*_{cs}V_{us}(\cP + \cPA)~,
\eea
where $\cP = \cP_s - \cP_d$ and $\cPA = \cPA_s - \cPA_d$. The weak phase
difference ($\phi$) between $V^*_{cs}V_{us}$ and $V^*_{cd}V_{ud}$ is a tiny
bit different from $\pi$:
\beq
\sin\phi~~=~~-~\frac{|V_{cb}||V_{ub}|}{|V_{cs}||V_{us}|}~\sin\g~~\sim~~
-~6.8\times10^{-4}~.
\eeq
However, the ratio of reduced penguin matrix elements to tree ones, in this
case, is at least an order of magnitude smaller:
\beq
\frac{|\cP + \cPA|}{|\cT_f|} \approx \frac{|P + PA|}{|T_f|} \sim 0.2
\eeq
Thus the associated CP asymmetry is also an order of magnitude smaller and we
can ignore this contribution.

\begin{figure}
\begin{center}
\includegraphics[width=0.45\textwidth]{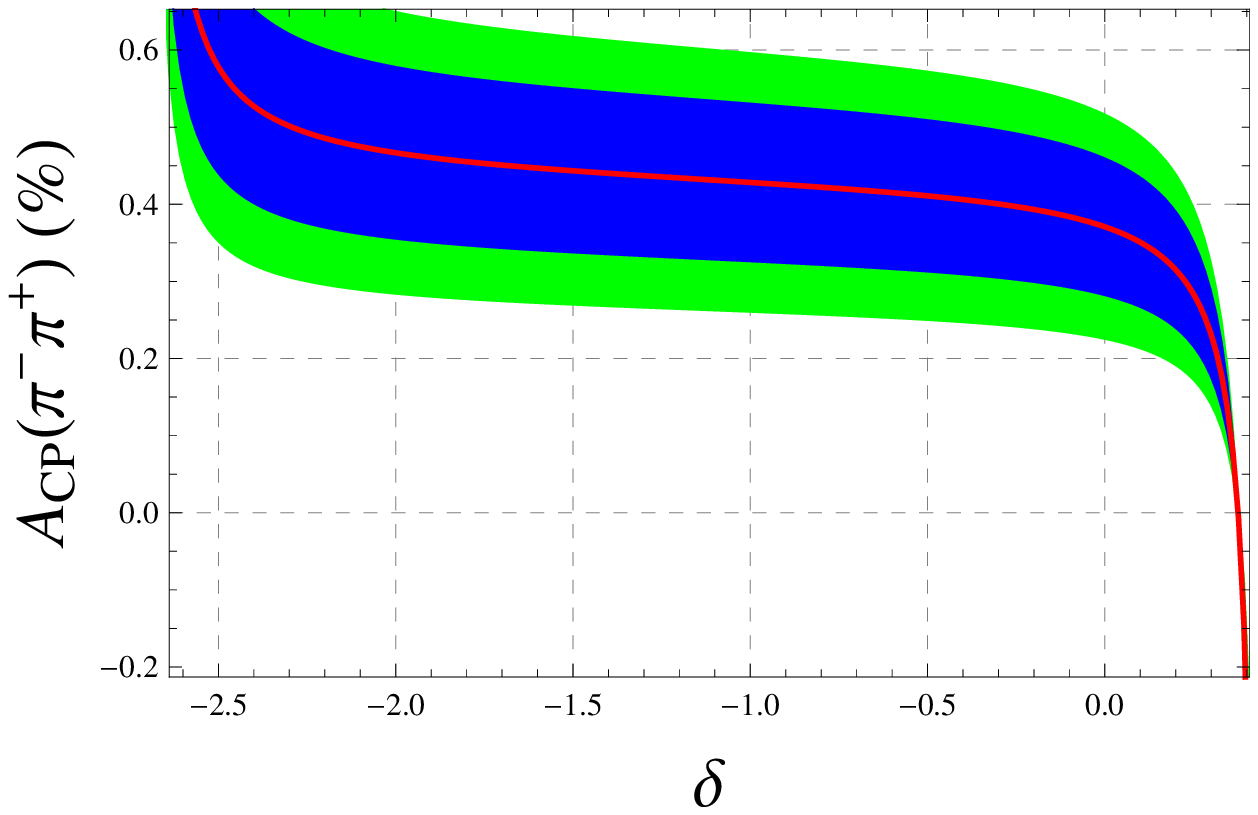} \hspace{0.5cm}
\includegraphics[width=0.45\textwidth]{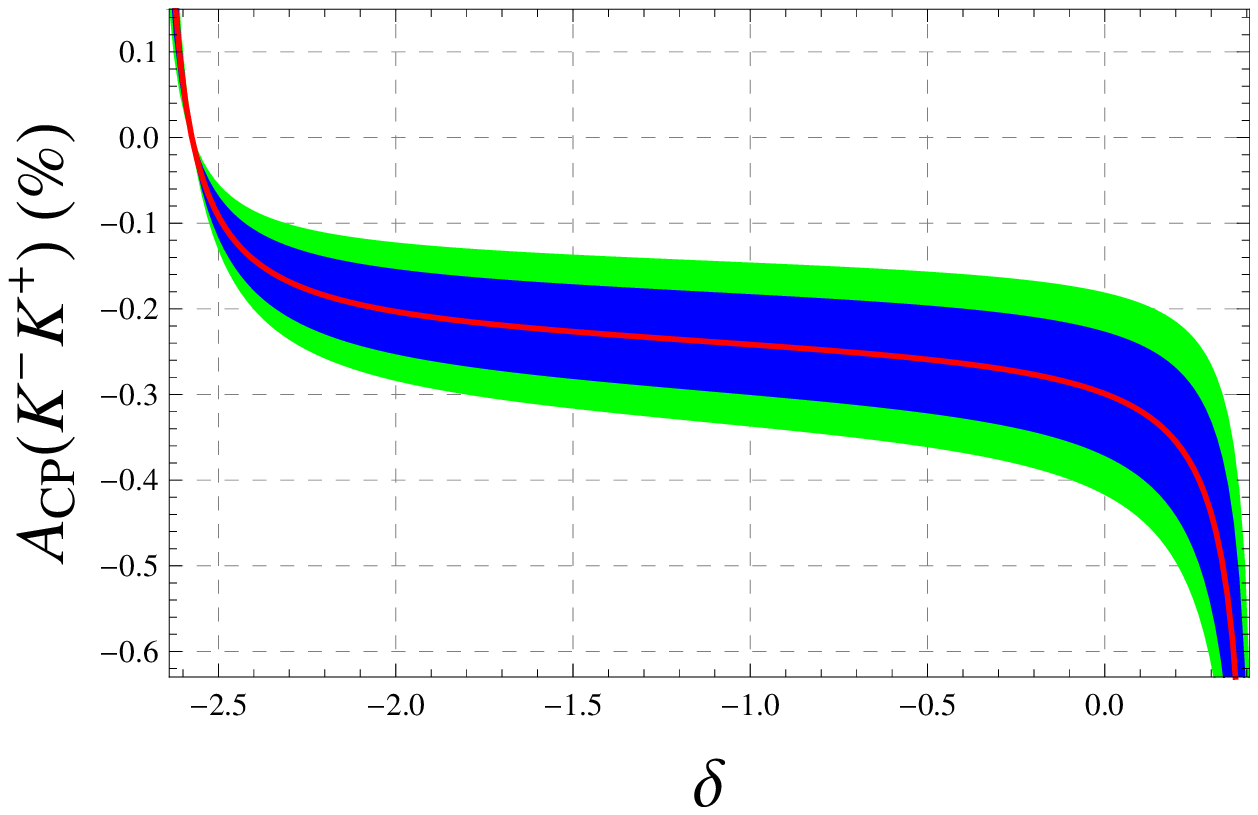}
\includegraphics[width=0.45\textwidth]{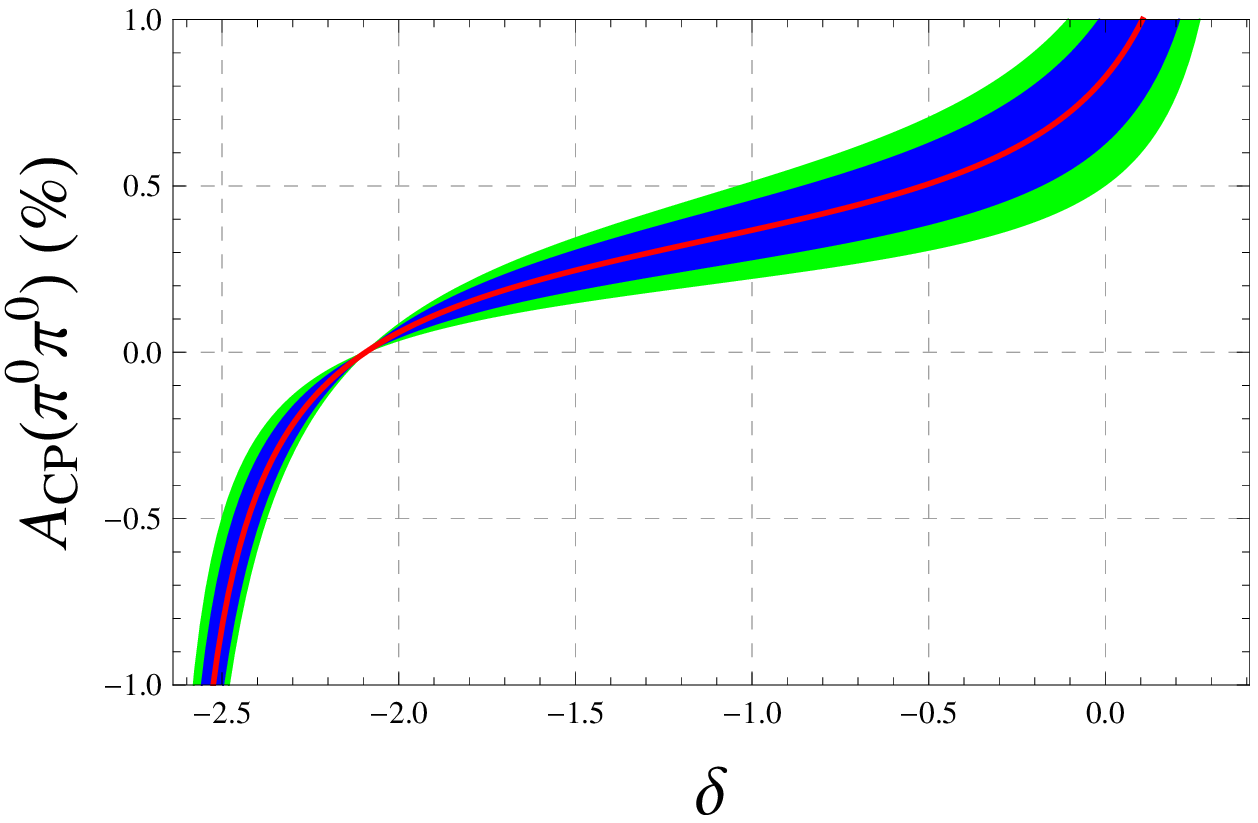} \hspace{0.5cm}
\includegraphics[width=0.45\textwidth]{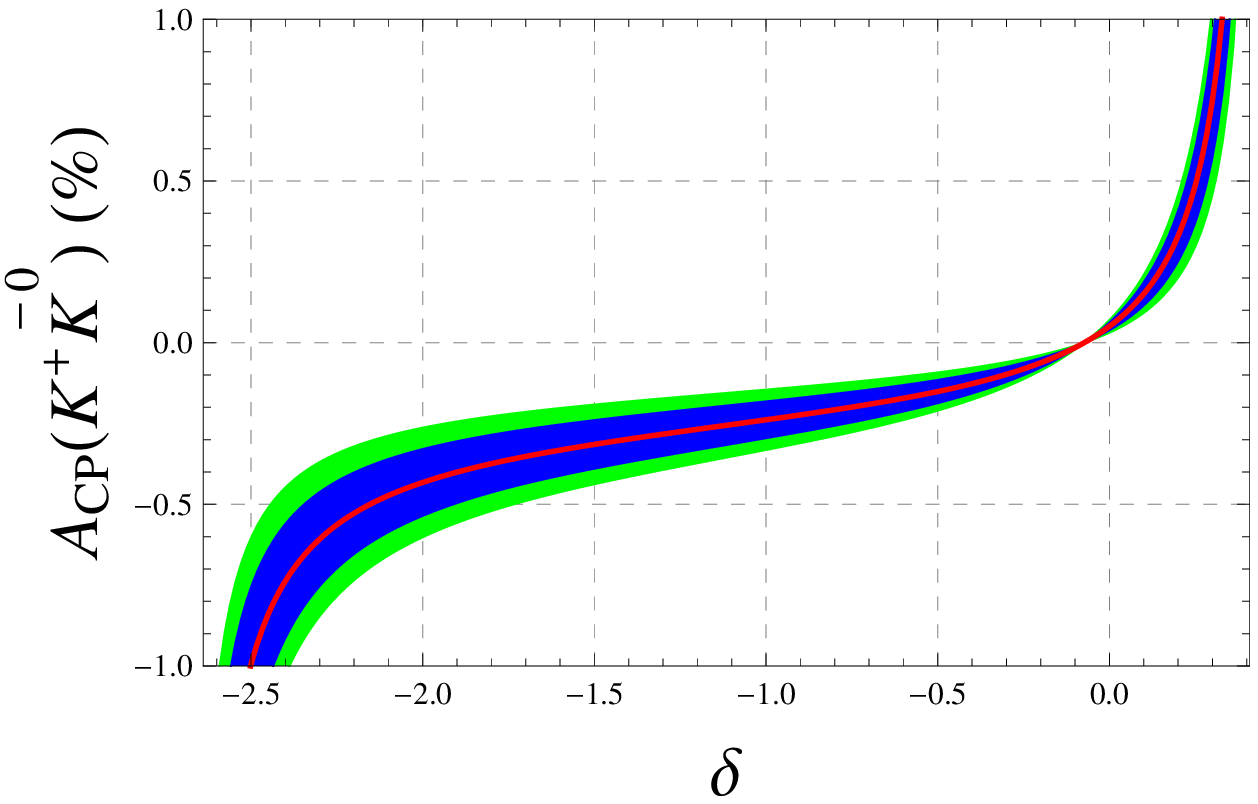}
\caption{Direct CP asymmetry ($A_{CP}(f)$) in various $D^0$ and $D^+$ decay modes
plotted as a function of the allowed values of $\delta$. The (red) lines represent
the central values, while inner (blue) and outer (green) bands respectively represent
68\% confidence level (1$\sigma$) and 90\% confidence level (1.64$\sigma$) regions
based on error in $\Delta A_{CP}$.
\label{fig:ACP}}
\end{center}
\end{figure}

Now that we have constrained $|P_b|$ as a function of $\delta$, we may use
it to explore direct CP asymmetries in other $D^0$ and $D^+$ processes. In Fig.\
\ref{fig:ACP}, we use Eq.\ (\ref{eqn:CPV}) and values of $\phi^f_T$ from Table
\ref{tab:SCS2P}, to plot the direct CP asymmetries in the processes $D^0\to(
\pi^+\pi^-, K^+K^-, \pi^0\pi^0)$ and $D^+\to K^+\ok$ as a function of the
strong phase $\delta$, within the range allowed by the CDF bounds.
The CP asymmetries for the final states $\pi^+\pi^-$ and $K^+K^-$ are found to
follow the central values of the CDF measurements quoted in Eq.\ (\ref{eqn:limits}).
The $\pi^+\pi^-$ CP asymmetry is found to be positive, while the $K^+K^-$ CP
asymmetry is found to be negative for a large range of allowed $\delta$ values.
In order to be able to pinpoint $\delta$, however, we need more precise
measurements of the individual asymmetries in these two channels. We also note
the correlation between the CP asymmetries in $\pi^0\pi^0$ and $K^+\ok$. The
former is found to be positive while the later is found to be negative, for a
large range of $\delta$. The best experimental value of the CP asymmetry in
$K^+\ok$, from the Belle experiment \cite{Ko:2010ng}, has large error bars and is
consistent with zero.

Notice that amplitudes not involving $P_b$ ($D^0\to K^0 \ok$ and $D^+\to\pi^+
\pi^0$) automatically have $A^{\rm dir}_{CP} = 0$. One can generate a non-zero
direct CP asymmetry in $D^0\to K^0\ok$ by including a $PA_b$ contribution, but
it is considerably harder to generate a large CP asymmetry in $D^+\to
\pi^+\pi^0$. Due to Bose statistics the $\pi^+\pi^0$ state has $I = 2$ and thus
can't get contributions from $\Delta I = 1/2$ penguin topologies. (There may
still be electroweak-penguin contributions. However these are expected to be too
small in $D$ decays to give rise to a percent level CP asymmetry.)

$D^+_s$ decay rates have large error bars, and measurements of CP asymmetries in
$D^+_s$ processes are as yet unavailable. The $D^+_s$ CP asymmetry discussion
has therefore been left out of the present analysis.

\section{Conclusions}

In this talk we explored the possibility of using flavor-SU(3) symmetry to study
branching ratios and direct CP asymmetries in $D^0$ and $D^+$ decays. Based on
our study of branching ratios we may arrive at the following conclusions:
\begin{itemize}
    \item The flavor-SU(3) framework works fairly well with CF $D$ decays. A
          $\chi^2$ minimization fit to extract the parameters yields a low value
          $\chi^2$ minimum.
    \item Measured branching fractions of SCS $D^0$ decays to $\pi^+\pi^-$ and
          $K^+K^-$ deviate considerably from flavor-SU(3) predictions; factorizable
          SU(3) breaking in ``Tree'' amplitudes does not provide a satisfactory
          explanation.
    \item We propose a model for SU(3) breaking in SCS $D^0\to PP$ decays, which is
          realized through an absence of GIM cancellation in penguin topologies.
    \item $D^0$ decays seem to follow the proposed SU(3) breaking scheme. However,
          it seems quite early to comment on $D^+$ and $D^+_s$ decays, where the
          error bars are quite large.
    \item $D\to PV$ decays are interesting, but these processes involve many
          more parameters and there are not enough data available.
    \item Flavor SU(3) is successful in explaining $I = 0$ dominance in the
          $D^0\to\pi^0\pi^+\pi^-$ Dalitz plot; cross ratios in several other
          Dalitz plots seem to agree with flavor-SU(3) predictions.
\end{itemize}

In the second part of the talk we applied our results from the study of branching ratios
in $D$ decays to explore the possibility of explaining recent CP violation measurements
in SCS $D$ decays and also predicting CP violation in other $D^0$ and $D^+$ decays. From
this discussion we may infer the following:
\begin{itemize}
    \item The recent LHCb and CDF $\Delta A_{CP}$ measurements can be explained by
          considering an enhanced CP-violating penguin, something that is not unusual to
          expect within the framework of the SM.
    \item Given the present experimental limits, the strong phase ($\delta$) of
          the CP-violating penguin amplitude is still fairly unconstrained.  In
          order to pinpoint $\delta$, a more precise determination of the
          individual CP asymmetries in the $\pi^+\pi^-$ and $K^+K^-$ final
          states is necessary. In turn this will lead to a better prediction of
          CP asymmetries in several other final states.
    \item We predict $A_{CP}$ in $D^+\to K^+\ok$ and $D^0\to\pi^0\pi^0$; these CP asymmetries
          seem to be correlated; $K^+\ok$ is negative while $\pi^0\pi^0$ is positive for a
          large range of $\delta$.
    \item In the model that we presented $A_{CP} = 0$ in $D^0\to K^0\ok$ and $D^+\to\pi^+\pi^0$;
          a non-zero $A_{CP}$ in the former case may be explained by introducing a CP-violating
          penguin topology of the annihilation type ($PA_b$).
    \item $A_{CP} \ne 0$ in $D^+\to\pi^+\pi^0$ needs new dynamics with both weak and strong phases
          different from the SM tree-level topologies.
\end{itemize}

In our present study we left out a wide variety of final states, due to lack of experimental data
as well as theoretical understanding. The SCS $D^0$ decays with an $\eta$ or an $\eta'$ in the final
state also call for OZI-suppressed singlet-exchange and -annihilation topologies, which were not
studied. Furthermore, CP asymmetries in $D$ meson decays to a pseudoscalar and a vector present a
whole new subject. With future measurements from LHCb and super-B factories we hope that our
understanding of the SM at low energies will get even better.

\section*{Acknowledgements}
B.\ B.\ is grateful to the organizers of Charm 2012 as well as D.\ London for arranging financial
support to attend the conference. This work was supported in part by the United States Department
of Energy under Grant No.\ DE FG02 90ER40560 and by NSERC of Canada. J.\ L.\ R.\ thanks the Aspen
Center for Physics for hospitality during the completion of this report. M.\ G.\ thanks the CERN
Theoretical Physics Unit for its kind hospitality during the completion of this report.

\end{document}